\documentclass[CEJP,PDF]{cej} 
\usepackage{layout}
\usepackage{amsmath}
\usepackage{textcomp}
\usepackage{hyperref}
\usepackage[inline]{trackchanges}

\newcommand{\barA}{{\overline{A}}}

\newcommand{\sigmarixs}{\sigma_{\mathrm{KH}}}

\newcommand{\bfP}{{\mathbf{P}}}

\newcommand{\bfz}{{\mathbf{z}}}
\newcommand{\bfa}{{\mathbf{a}}}
\newcommand{\bfb}{{\mathbf{b}}}

\newcommand{\ee}{{\mathrm{e}}}

\newcommand{\bfk}{{\mathbf{k}}}

\newcommand{\bfr}{{\mathbf{r}}}

\newcommand{\bfL}{{\mathbf{L}}}

\newcommand{\ninej}[9]{\left\{ \begin{array}{@{}c@{\;}c@{\;}c@{}}
         #1 & #2 & #3 \\
         #4 & #5 & #6 \\
         #7 & #8 & #9
         \end{array}\right\}}
\newcommand{\sixj}[6]{\left\{ \begin{array}{@{}c@{\;}c@{\;}c@{}}
         #1 & #2 & #3 \\
         #4 & #5 & #6
         \end{array}\right\}}
\title{Angular dependence of resonant inelastic x-ray scattering: a spherical tensor expansion}

\articletype{Research Article}

\author{Am\'elie Juhin\inst{1}\email{Amelie.Juhin@impmc.upmc.fr},
        Christian Brouder\inst{1}\email{Christian.Brouder@impmc.upmc.fr},
        Frank de Groot\inst{2}\email{F.M.F.deGroot@uu.nl}}

\institute{ 
     \inst{1} Institut de Min\'eralogie, de Physique des Mat\'eriaux et de Cosmochimie, Sorbonne Universit\'es,
UMR CNRS 7590, UPMC Univ. Paris 06, Mus\'eum National d'Histoire Naturelle, IRD UMR 206,
4 place Jussieu, F-75005 Paris, France.
     \inst{2} Inorganic Chemistry and Catalysis,~Department of
Chemistry, Utrecht University, Universiteitsweg 99, 3584 CG Utrecht, 
The Netherlands
          }

\abstract{A spherical tensor expansion is carried out to express the resonant inelastic scattering cross-section as a
sum of products of fundamental spectra with tensors involving wavevectors 
and polarization vectors of incident and scattered photons. 
The expression presented in this paper differs from that of the 
influential article by Carra et al. (Phys. Rev. Lett. 74, 3700, 1995) 
because it does not omit interference terms between electric dipole and 
quadrupole contributions when coupling each photon to itself. 
Some specific cases of the spherical tensor expansion are discussed. 
For example the case of isotropic samples is considered
and the cross-section is expressed as a combination of only three 
fundamental spectra for the situation where electric dipole or 
electric quadrupole transitions in the absorption process are followed by electric dipole transitions in the emission. This situation includes the case of untextured powder samples, which corresponds to the most frequent situation met experimentally. Finally, it is predicted that some circular dichroism may be observed on isotropic samples provided that the circular polarization of the scattered beam can be detected.}

\keywords{Resonant Inelastic Scattering, angular dependence, dichroism}
\pacs{61.05.cf, 78.70.Ck}

\begin{document}
\maketitle


\section{Introduction}

Resonant inelastic scattering (RIS) spectroscopies are remarkable tools
to study electronic, magnetic and vibrational properties of
materials \cite{Ament-11}. They span a broad energy range,
from infrared frequencies for phonon excitations, through
optical photons for electronic excitations to x-ray
energies for Resonant Inelastic X-ray Scattering (RIXS). 
The richness of these spectroscopies is due to the
large number of possible spectra obtained by varying the
energy, direction and polarization state of the incident and scattered
electromagnetic waves.
As a matter of fact, there is so much information
in the spectra that it is difficult to know whether
a specific set of experiments measures all potential information.
The main purpose of this paper is to
determine a finite set of fundamental spectra in terms of
which all possible experimental spectra can be expressed.
More precisely, the resonant inelastic scattering spectrum obtained for a
given wavevector and polarization vector of the incident
beam ($\bfk$ and $\epsilon$) and
scattered beam ($\bfk_s$ and $\epsilon_s$)
is written as a sum of terms which are
fundamental
spectra~\cite{Brouder-93-Schille,Thole-94-2} (independent of the
incident and scattered beam) multiplied by
an explicit polynomial in
$\bfk$, $\epsilon$, $\bfk_s$ and $\epsilon_s$.
 
The fundamental spectra will be computed by a
spherical tensor analysis, which was used with great success 
for the x-ray photoemission of localized magnetic
systems~\cite{Thole-91,Laan-93,Thole-94,vanderlaan-1995a,vanderlaan-1995b} and in x-ray absorption 
spectroscopy \cite{Thole92,Carra93,Brouder-93-Schille,Thole-94-2,Brouder}.
For the case of x-ray absorption~\cite{Brouder}
such geometric (\textit{i.e.} coordinateless) and fully-decoupled expressions
are useful (i) to disentangle the properties of the sample
from those of the measurement; (ii) to determine
specific experimental arrangements aiming at 
the observation of specific sample properties; (iii)
to provide the most convenient starting point 
to investigate the reduction of the number
of fundamental spectra due to crystal symmetries.

Following the same idea in the present case, the resonant inelastic
scattering cross-section is expressed as the smallest possible 
combination of fundamental spectra. To do so, the remarkable angular momentum recoupling techniques developed by the Lithuanian school \cite{Jucys, Rudzikas}
are used.
As an application of the general results,
the most common experimental case of an isotropic sample (disordered
molecules, liquids, polycrystals or powders) is described in
detail. It is demonstrated that the spectrum of an isotropic sample 
for electric dipole excitation 
and electric dipole emission with undetected scattered polarization is 
the sum of only three fundamental spectra, compared to 19 in the general 
spherical-tensor-based expression (and versus the 81 components of a
general fourth-rank Cartesian tensor). 

The starting point of this work is the
Kramers-Heisenberg formula \cite{Kramers-25,Kramers-25-GB}, extended to take electric quadrupole transitions into account.
The Thomson scattering term is not explicitly considered but it can be easily included (see below). For simplicity, the non-resonant term in the Kramers-Heisenberg formula is neglected (i.e., assuming the vicinity of an absorption edge). If the non-resonant term is sizeable~\cite{Blume}, it can be taken into account by similar methods. 
It would also be possible to take
variable decay lifetimes into account \cite{Wray-12}, but
in this paper only the standard Kramers-Heisenberg formulation is considered
for simplicity. Electric dipole and quadrupole transitions contribute
significantly in the
x-ray range \cite{BrouderAXAS}. Using x-ray photons, typically from a synchrotron radiation source, one can
choose a specific atomic species and orbital in 
a complex compound by selecting the suitable absorption edge. The sample can be magnetic, it can be submitted to an
external electric or magnetic field, as long as the orientation
of the external fields remains constant with respect to
the sample. Thus, the results presented in this paper are a 
pure group theoretical consequence of 
the resonant scattering cross-section formula. 
Such an expansion does not need to assume 
that a specific edge is measured (\textit{i.e.}, only the nature of the 
transitions involved is specified) nor to 
assume that the states involved in the transitions are localized 
or delocalized. 

The paper consists of three parts. In the first part, the spherical tensor expression is derived
for the resonant scattering amplitude. Similar expressions were already 
published~\cite{Veenendaal-98,Marri-PhD,MarriCarra,Marri-06}. 
The second part of this paper consists of a full recoupling
of the scattering amplitude to obtain the spherical tensor 
decomposition of the scattering cross-section. At this stage, several works in the literature make
specific approximations, for example by considering a localized initial
state or by using the fast collision approximation 
(which is not generally valid~\cite{Ferriani-04}).
Finally,
recoupling techniques are used to separate the incident beam
from the scattered beam in the cross section and to separate
the polarization part from the wavevector part.
This enables to treat the frequent experimental 
case where the polarization state of the scattered beam is not measured. 

Related results were obtained in specific experimental 
conditions \cite{Braicovich-1996,Braicovich-1999} or in specific coordinate
systems \cite{Fukui-01,Fukui-01-2,Ogasawara-04} but this
general expression is new. Its coordinateless form enables
descriptions of the general form of the resonant inelastic
spectroscopy of isotropic samples. 
Other works use an approach similar 
to that presented in this paper \cite{Carra-95,Veenendaal-98,Ament-11},
but they do not take into account the
interference between electric dipole and electric quadrupole
transitions which generates natural
circular dichroism \cite{Alagna-98,BrouderNCD}, 
or they carry out a different coupling.

A consequence of the formula presented in this paper is that 
circular dichroism can be observed
for isotropic (in particular not magnetically oriented) samples,
if the polarization of the scattered beam can be measured experimentally.
The third part of the paper contains appendices giving 
the detail of the derivations.

\section{General case}
\subsection{The Kramers-Heisenberg formula}
\label{KHformula-sect}
The scattering of light by a quantum system is described
by an equation derived by Kramers and Heisenberg before
the advent of quantum theory \cite{Kramers-25}. Its first
quantum derivation (in the electric-dipole approximation)
is due to Dirac \cite{Dirac-27-dispersion}.
The multipole scattering cross-section is~\cite{Tulkki}:

\begin{eqnarray}
\sigma_\mathrm{SCAT}
= r_e^2
   \frac{\omega_{s}}{\omega}\sum_F
\Big|
 \epsilon_s^*\cdot\epsilon
\langle F |\ee^{i(\bfk-\bfk_s)\cdot\bfr}|I\rangle
+ \frac{1}{m}\sum_N
   \frac{\langle F| \epsilon_s^*\cdot\bfP
   \ee^{-i\bfk_s\cdot\bfr} |N\rangle
\langle N| \epsilon\cdot\bfP
   \ee^{i\bfk\cdot\bfr} |I\rangle}
   {E_I-E_N+\hbar\omega+i\gamma}
\nonumber\\
+ \frac{1}{m}\sum_N
   \frac{\langle F| \epsilon\cdot\bfP
   \ee^{i\bfk\cdot\bfr} |N\rangle
\langle N| \epsilon_s^*\cdot\bfP
   \ee^{-i\bfk_s\cdot\bfr} |I\rangle}
   {E_I-E_N-\hbar\omega_{s}+i\gamma}
   \Big|^2 
\delta(E_F+\hbar\omega_{s}-E_I-\hbar\omega)
\label{scatcca}
\end{eqnarray}
where
$m$ is the electron mass, $r_e$ is the classical electron radius: 
$r_e=e^2/(4\pi\epsilon_0 mc^2)$,
$|I\rangle$, $|N\rangle$, $|F\rangle$ are respectively the initial, intermediate and final states,
$\gamma$ is the total width of the intermediate state
$|N\rangle$, $\bfP$ and $\bfr$ are the momentum and position operators. The incident and scattered photons are characterized by the pulsation, 
wavevector and polarization vectors $\omega,\bfk,\epsilon$ and
$\omega_s,\bfk_s,\epsilon_s$, respectively. Note that $\epsilon_s^*$ 
denotes the complex conjugate of $\epsilon_s$. 

The first term of this expression describes Thomson scattering, which will not be considered explicitly here. 
$E_I$ being negative and large, $E_I+\hbar\omega$ can be small, 
and $E_I-\hbar\omega_s$ is large. Hence, it can 
generally be assumed that the 
third matrix element in Equation\eqref{scatcca} can be neglected 
with respect to the second one (although it could be
treated with similar methods). Therefore, only the second 
transition amplitude remains in the expression of the scattering 
cross-section, yielding the well-known partial Kramers-Heisenberg formula: 

\begin{eqnarray}
 \sigmarixs
 = \frac{r_e^2}{m^2}
   \frac{\omega_{s}}{\omega}\sum_F
\Big|
\sum_N
   \frac{\langle F| \epsilon_s^*\cdot\bfP
   \ee^{-i\bfk_s\cdot\bfr} |N\rangle
\langle N| \epsilon\cdot\bfP
   \ee^{i\bfk\cdot\bfr} |I\rangle}
   {E_I-E_N+\hbar\omega+i\gamma}
   \Big|^2 
   \delta(E_F+\hbar\omega_{s}-E_I-\hbar\omega).
\label{scatcc}
\end{eqnarray}

For notational convenience, a single variable
$\bfP \ee^{i\bfk\cdot\bfr}$ is written
instead of a sum over all electrons of the system
$\sum_{j=1}^N \bfP_j \ee^{i\bfk\cdot\bfr_j}$. 

\subsection{Multipole expansion}
\label{mulpolsect}
First the matrix element $\langle N| \epsilon\cdot\bfP
\ee^{i\bfk\cdot\bfr} |I\rangle$,
describing the absorption from the initial state $|I\rangle$ 
to the intermediate state $|N\rangle$, is transformed by 
expanding $\ee^{i\bfk\cdot\bfr}$ to first order:
$\ee^{i\bfk\cdot\bfr}\simeq 1+i\bfk\cdot\bfr$.
Hence,
\begin{eqnarray*}
\langle N| \epsilon\cdot\bfP \ee^{i\bfk\cdot\bfr} |I\rangle
&\simeq& 
\langle N| \epsilon\cdot\bfP |I\rangle
+
i
\langle N| \bfk\cdot\bfr \epsilon\cdot\bfP |I\rangle.
\end{eqnarray*}
The electric dipole matrix element is transformed by using the equation of motion of $\bfP$ which is
$\bfP=(m/i\hbar) {[}\bfr,H_0{]}$~\cite{BrouderAXAS}. Thus,
\begin{eqnarray*}
\langle N| \epsilon\cdot\bfP |I\rangle
&=&(m/i\hbar) (E_I-E_N) \langle N| \epsilon\cdot\bfr |I\rangle.
\end{eqnarray*}
For the quadrupole matrix element, one uses the identity from 
Ref.~\cite{Blinder},
\begin{eqnarray*}
\bfk\cdot\bfr \epsilon\cdot\bfP &=&
-(im/2\hbar) {[}\epsilon\cdot\bfr \bfk\cdot\bfr,H_0{]}
+1/2 (\bfk\times\epsilon)\cdot\bfL,
\end{eqnarray*}
where $\bfL$ is the angular momentum operator. 
In this work the second (magnetic dipole) term
is not taken into account because it is small
in the x-ray range~\cite{BrouderAXAS}. Therefore
\begin{eqnarray*}
\langle N| \epsilon\cdot\bfP \ee^{i\bfk\cdot\bfr} |I\rangle
&\simeq& 
-\frac{im}{\hbar} (E_I-E_N) 
\Big(\langle N| \epsilon\cdot\bfr |I\rangle
+\frac{i}{2} \langle N| \epsilon\cdot\bfr \bfk\cdot\bfr|I\rangle\Big)
\\&=& -\frac{im}{\hbar}(E_I-E_N)\sum_{\ell=0}^{1}f_{\ell}\langle N|\epsilon\cdot\bfr (\bfk\cdot\bfr)^{\ell}|I\rangle,
\end{eqnarray*}
with $f_0=1$ and $f_1=\frac{i}{2}$.\\

Similarly, on transforming the matrix element describing the emission from the intermediate state $|N\rangle$ to the final state $|F\rangle$, 
\begin{eqnarray*}
 \langle F| \epsilon_s^*\cdot\bfP \ee^{-i\bfk_s\cdot\bfr} |N\rangle
&\simeq& 
-\frac{im}{\hbar} (E_N-E_F)
\Big(
\langle F| \epsilon_s^*\cdot\bfr |N\rangle
-\frac{i}{2} \langle F| \epsilon_s^*\cdot\bfr \bfk_s\cdot\bfr|N\rangle\Big)
\\&=& -\frac{im}{\hbar}(E_N-E_F)\sum_{{\ell}'=0}^{1}f_{\ell'}^*\langle F|\epsilon_s^*\cdot\bfr (\bfk_s\cdot\bfr)^{\ell'}|N\rangle.
\end{eqnarray*}

Finally,
\begin{eqnarray}
\nonumber
\sigmarixs
 = \frac{r_e^2}{\hbar^2}\frac{\omega_{s}}{\omega}\sum_F
\Big|
\sum_N \frac{(E_I-E_N)(E_N-E_F)}{E_I-E_N+\hbar\omega+i\gamma} 
\sum_{{\ell},{\ell'}=0}^{1}f_{\ell}f_{\ell'}^*
 \langle N|\epsilon\cdot\bfr (\bfk\cdot\bfr)^{\ell}|I\rangle \langle F|\epsilon_s^*\cdot\bfr (\bfk_s\cdot\bfr)^{\ell'}|N\rangle
   \Big|^2 \delta_{E}, 
   \label{scatcc2}
\end{eqnarray}
where $\delta_{E}=\delta(E_F+\hbar\omega_{s}-E_I-\hbar\omega)$. 
Denoting
$C=r_e^2\omega_s/(\hbar^2\omega)$ and
\begin{eqnarray*}
F_{I,N,F}&=&\sum_{{\ell},{\ell'}=0}^{1}f_{\ell}f_{\ell'}^*
 \langle N|\epsilon\cdot\bfr (\bfk\cdot\bfr)^{\ell}|I\rangle \langle
F|\epsilon_s^*\cdot\bfr (\bfk_s\cdot\bfr)^{\ell'}|N\rangle,
\end{eqnarray*}
Equation\eqref{scatcc2} becomes:
\begin{eqnarray}
\sigmarixs
 &=& C \sum_F
\Big|
\sum_N \frac{(E_I-E_N)(E_N-E_F)}{E_I-E_N+\hbar\omega+i\gamma}F_{I,N,F}
   \Big|^2 \delta_{E}.
   \label{scatcc3}
\end{eqnarray}

Equation\eqref{scatcc3} is a general expression describing the 
resonant inelastic scattering intensity for any combination of 
the absorption and emission transition
operators. Each transition operator can be either pure electric 
dipole ($E_1$), or pure
electric quadrupole ($E_2$), or a mixture of both ($E_1+E_2$). 

\subsection{The Kramers-Heisenberg formula expressed in terms 
of spherical tensors}
The expression of the $F_{I,N,F}$ intensity factor 
appearing in the Kramers-Heisenberg equation (Equation\eqref{scatcc3}) is 
transformed using spherical tensors and their coupling
properties. This transformation is detailed in 
Appendices~\ref{exp} and \ref{couple}.
For a short introduction to spherical tensors and their
application to x-ray spectroscopies, the reader is referred 
to Ref.~\cite{Brouder} for the case of the X-ray absorption cross-section. 
First the notation is briefly explained. An $\ell$th-rank spherical 
tensor $T$ is written as $T^{(\ell)}$, not to be mistaken for
$T^{\ell}$, the ${\ell}$th power of 
$T$. Cartesian vectors, such as $\epsilon$, $\bfr$ or $\bfk$ are 
written in their usual form, \textit{i.e.}, without brackets. However one should keep in mind that 
Cartesian vectors correspond to
first-rank spherical tensors, and as such they shall also be 
written as 
$\epsilon^{(1)}$, $\bfr^{(1)}$, $\bfk^{(1)}$ or
$\epsilon^1$, $\bfr^1$, $\bfk^1$.

After the transformation of $F_{I,N,F}$ 
(see Appendices~\ref{exp}
and \ref{couple}), Equation\eqref{scatcc3} becomes:
\begin{eqnarray}
\label{eq4}
\sigmarixs \nonumber &=& C \sum_F
\Big|
\sum_N \sum_{g,{\ell},{\ell'}}\frac{(E_I-E_N)(E_N-E_F)}{E_I-E_N+\hbar\omega+i\gamma}
\frac{(-1)^g h_{\ell}h_{\ell'}^*}{\sqrt{(2\ell+3)(2\ell'+3)}}  
\nonumber\\&&
\Big\{ \{\epsilon_s^* \otimes \bfk_s^{\ell'} \}^{(\ell'+1)} \otimes \{\epsilon \otimes \bfk^{\ell} \}^{(\ell+1)}\Big\}^{(g)}
\cdot \Big\{\bfr_{FN}^{(\ell'+1)} \otimes  \bfr_{NI}^{(\ell+1)}\Big\}^{(g)}\Big|^2 \delta_{E},
\end{eqnarray}
where $g$ runs from $|\ell-\ell'|$ to $(\ell+\ell'+2)$, $\ell$ and $\ell'$ run from 0 to 1.  
The $h_{\ell}$ factors are defined by 
$h_0=-\sqrt{3}$, $h_1=\frac{i}{2}\sqrt{5}$.

In Refs.~\cite{Carra-95,Veenendaal-98}, a similar formula was obtained
in terms of vector spherical harmonics. 
The present coupling is chosen (as in \cite{Veenendaal-98}) 
to avoid irrelevant powers of $\sqrt{4\pi}$ in the final result.

The first tensor product $\Big\{ \{\epsilon_s^* \otimes \bfk_s^{\ell'} \}^{(\ell'+1)} \otimes \{\epsilon \otimes \bfk^{\ell} \}^{(\ell+1)}\Big\}^{(g)}$ 
characterizes the incident beam ($\epsilon, \bfk$) and the scattered beam 
($\epsilon_s, \bfk_s$). The variables describing the sample are gathered 
in the second tensor product $\Big\{\bfr_{FN}^{(\ell'+1)} \otimes  \bfr_{NI}^{(\ell+1)}\Big\}^{(g)}$.

Then, defining 
\begin{eqnarray}
A_{FI}^{(g)}(\ell, \ell')&=& \sum_N \frac{(E_I-E_N)(E_N-E_F)}{E_I-E_N+\hbar\omega+i\gamma} \{\bfr_{FN}^{(\ell'+1)} \otimes  \bfr_{NI}^{(\ell+1)}\}^{(g)},
\end{eqnarray}
one obtains
\begin{eqnarray}
\sigmarixs  = C \sum_F
\Big| \sum_{g,{\ell},{\ell'}}\frac{(-1)^g h_{\ell}h_{\ell'}^*}{\sqrt{(2\ell+3)(2\ell'+3)}} 
\Big\{ \{\epsilon_s^* \otimes \bfk_s^{\ell'} \}^{(\ell'+1)} \otimes \{\epsilon \otimes \bfk^{\ell} \}^{(\ell+1)}\Big\}^{(g)}
\cdot A_{FI}^{(g)}(\ell, \ell') \Big|^2  \delta_{E}.
\label{sc6}
\end{eqnarray}

Note that Thomson scattering can be taken into account as a contribution to the term $A_{FI}^{(0)}(0,0)$. 
In the case of x-rays, at this stage of the spherical tensor expansion, 
the local nature of the initial core orbital is 
often used in the literature to specify the
absorption edge~\cite{Veenendaal-96}.
Such an approach is powerful to derive sum rules, for instance.
In this paper a general initial state is used.

As remarked by Carra and coll. \cite{Carra-95}, it is much more
convenient to work with each photon coupled to itself 
instead of with the incident photon coupled to the scattered
one as in Eq.~(\ref{sc6}). This is achieved by expanding the
square modulus in Eq.~(\ref{sc6}) and recoupling the
spherical tensors describing the incident and scattered beams.
In the influential Ref.~\cite{Carra-95} the authors note that 
``This is a rather technical part in our derivation, and will not be
discussed here.''
The calculation is indeed lengthy and a detailed
derivation is given in Appendices~\ref{square1} and \ref{square2}). 
This allows the derivation of the final expression for $\sigmarixs$,
which now has fully decoupled sample- and 
beam-dependent parts:
\begin{eqnarray}
\nonumber \sigmarixs = \sum_{g_1,g_2,{\ell}_1,{\ell}_2, \ell'_1,\ell'_2}
\sum_{a,b,c,u,u',v,v'}(-1)^{a+\ell_2+\ell'_2-g_2} 
h_{\ell_1}h_{\ell'_1}^*
h_{\ell_2}^*h_{\ell'_2} \Pi_{g_1,g_2,b,c,u,v,u',v'} 
\nonumber\\ 
\ninej{\ell_1'+1~}{\ell_1+1~}{g_1}{\ell_2'+1~}{\ell_2+1~}{g_2}{b}{c}{a}
\ninej{1~}{\ell_1~}{\ell_1+1}{1~}{\ell_2~}{\ell_2+1}{u}{v}{c}
\ninej{1~}{\ell'_1~}{\ell'_1+1}{1~}{\ell'_2~}{\ell'_2+1}{u'}{v'}{b}
\gamma^{bca}_{UL}\cdot S^{g_1g_2a}_L, 
\label{KHfinal}
\end{eqnarray}
where $h_0=-\sqrt{3}$, $h_1=\frac{i}{2}\sqrt{5}$,
$U=(u,v,u',v')$ and
$L=(\ell_1,\ell_2,\ell'_1,\ell'_2)$.
The tensors $\gamma^{bca}_{UL}$ describe the
incident and scattered x-rays, the tensor
$S^{g_1g_2a}_L$ describes the sample. More precisely,
\begin{eqnarray}
\gamma^{bca}_{UL} &=& 
\Big \{ \mathrm{Out}_{UL}^{(b)} \otimes 
     \mathrm{In}_{UL}^{(c)} \Big \}^{(a)} ,
\end{eqnarray}
is obtained by coupling the tensors
$\mathrm{In}_{UL}^{(c)}$ of the incident beam 
and the tensors $\mathrm{Out}_{UL}^{(b)}$
of the scattered (outgoing) beam, where
\begin{eqnarray}
\mathrm{Out}_{UL}^{(b)} &=& \Big \{ \{ \epsilon_s^* \otimes 
\epsilon_s  \}^{(u')} \otimes \{ \bfk_s^{\ell'_1} \otimes 
\bfk_s^{\ell'_2} \} ^{(v')}\Big\}^{(b)},
\label{Out}\\
\mathrm{In}_{UL}^{(c)} &=& 
\Big \{ \{ \epsilon \otimes \epsilon^*  \}^{(u)} \otimes 
\{ \bfk^{\ell_1} \otimes \bfk^{\ell_2} \} ^{(v)}\Big\}^{(c)}.
\label{In}
\end{eqnarray}

The tensors describing the sample are
\begin{eqnarray}
S^{g_1g_2a}_L = \frac{r_e^2\omega_s}{\hbar^2\omega} \sum_F
 \Big \{ A_{FI}^{(g_1)}(\ell_1, \ell'_1) \otimes  \barA_{IF}^{(g_2)}
  (\ell_2, \ell'_2)  \Big \} ^{(a)}
  \delta(E_F+\hbar\omega_s -E_I-\hbar\omega),
\end{eqnarray}
where
\begin{eqnarray*}
A_{FI}^{(g)}(\ell, \ell')&=& \sum_N \frac{(E_I-E_N)(E_N-E_F)}
    {E_I-E_N+\hbar\omega+i\gamma} 
  \{\langle F|\bfr^{({\ell'}+1)}|N\rangle  \otimes  
  \langle N|\bfr^{({\ell}+1)}|I\rangle\}^{(g)},\\
 \barA_{IF}^{(g)}(\ell, \ell')&=& \sum_N \frac{(E_I-E_N)(E_N-E_F)}
    {E_I-E_N+\hbar\omega-i\gamma} 
  \{\langle I|\bfr^{({\ell}+1)}|N\rangle  \otimes  
  \langle N|\bfr^{({\ell'}+1)}|F\rangle\}^{(g)}.
\end{eqnarray*}

The spherical tensors $S^{g_1g_2a}_L$ are the fundamental spectra
from which all experimental spectra can be expressed.
These fundamental spectra are weighted by coefficients 
$\gamma^{bca}_{UL}$ which are polynomials
in $\epsilon$, $\epsilon_s$, $\bfk$ and $\bfk_s$ and which describe the
experimental conditions.
The three $9j$-factors (and the related triangular conditions) ensure that 
these variables are coupled in the scattering process in a 
correct (physical) way.

The result presented in this paper differs from that of Ref.~\cite{Carra-95} because, as noted by Ferriani~\cite{Ferriani-PhD}, these authors
neglect some interference effects. More precisely, they replace the square 
modulus of the sum
$|\sum_{\ell\ell'}\dots |^2$ in eq.~(4) of the present paper
by the sum of the square moduli $\sum_{\ell\ell'}|\dots |^2$.
Since this approximation is not made here, there is an additional $9j$ 
recoupling coefficient describing the interference between 
electric dipole and quadrupole
contributions when each photon is coupled to itself. 
These interference terms are physically relevant since, 
even in the simpler case of x-ray absorption, they lead to important effects
such as natural circular dichroism~\cite{Alagna-98,BrouderNCD},
non-reciprocal gyrotropy~\cite{Goulon} and
magnetochiral dichroism~\cite{Goulon-02}.

The sum rules derived in Ref.~\cite{Carra-95,Veenendaal-96}
may have a restricted range of validity inasmuch as they neglect these interference terms. 
In subsequent articles \cite{Veenendaal-98,Ament-11},
the multipole matrix elements are coupled in a different
way, which avoids the additional $9j$-coefficients but
is no longer compatible with sum rules or Green function
representations.
  
Applications of the above general expression to particular cases will be 
given as examples in Section \ref{particular}, but before that
the range of values of the different variables
in Eq.~\eqref{KHfinal} is specified.
  
\subsection{Possible values of the different indices appearing
in Equation\eqref{KHfinal}}
The ``selection rules'' of the 9-$j$ symbols give the following
possible values taken by the angular variables  in
Equation\eqref{KHfinal}.
\begin{itemize}
	\item {$0 \leq \ell_i \leq 1$ (for $i=1,2$ ): these indices deal with the transition operator in the absorption. \\$\ell_i=0$~ for~dipole~excitation,~$\ell_i=1$~for~quadrupole excitation}
	\item {$ 0 \leq \ell_i' \leq 1$  (for $i=1,2$ ): these indices deal with the transition operator in the emission.\\ $\ell_i'=0$~ for~dipole~emission,~$\ell_i'=1$~for~quadrupole emission}
	\item { $|\ell_i -\ell_i'| \leq g_i \leq \ell_i + \ell_i' +2$  ( for $i=1,2$ ) : these indices couple absorption and emission transition operators.}
	\item { $0 \leq u \leq  2$: $u$ deals with the polarization of the incident beam.}
	\item { $0 \leq u' \leq 2$: : $u'$ deals with the polarization of the scattered beam.}
  \item { $|\ell_1-\ell_2| \leq v \leq  \ell_1+\ell_2$: $v$ deals with the direction of the incident beam.}
  \item { $|u-v| \leq c \leq  u+v$ and $|\ell_1 -\ell_2| \leq c \leq \ell_1 + \ell_2 +2 $: $c$ gathers all characteristics of the incident beam. }
  \item { $|\ell'_1-\ell'_2| \leq v' \leq \ell'_1+\ell'_2$: $v'$ deals with the direction of the scattered beam. }
  \item { $|u'-v'| \leq b \leq  u'+v'$ and $|\ell_1' -\ell_2'| \leq b \leq \ell_1' + \ell_2'+2$: $b$ gathers all characteristics of the outgoing beam. }
  \item { $|b-c| \leq a \leq b+c $ and $|g_1 -g_2| \leq a \leq g_1+g_2$ : $a$ couples everything. }
\end{itemize}

\section{Applications to special cases}
\label{particular}
In this section several special cases of Equation\eqref{KHfinal}
are considered.

\subsection{Conditions due to the type of polarization}
Knowing the type of beam polarization reduces the number of values that 
the indices $u$ and $u'$ are allowed to take. 
When the incident beam is linearly polarized, then $u=0$ or $2$. 
Note that there is a basic difference between the polarization
of the incident beam and that of the scattered beam.
Indeed, the incident beam is prepared by the experimental setup
($\epsilon$ can be tuned as one wishes) while the polarization
state of the scattered beam is entirely determined by the
incident beam and the sample. It cannot be tuned.
However, the polarization properties of
the scattered beam can be measured. 
If the requirement is to measure the intensity
along the polarization direction $\epsilon_s$, then
that $\epsilon_s$ is introduced in the formula. If 
the polarization state is not measured, then the trace of the
density matrix representing the polarization state of the
scattered beam can be taken.
In other words, the measured cross-section is equal to the 
sum over two perpendicular directions of the scattered 
polarization (see Appendix \ref{average}). In this case the term
corresponding to $u'=1$ in Equation\eqref{KHfinal} vanishes
since $\langle \{\epsilon_s\otimes\epsilon_s^*\}^{(1)} \rangle =0$.

It is important to discuss the case where the polarization of
the scattered beam is not measured. As proved in 
Appendix~\ref{average}, this corresponds to an average
of polarizations but not to an ``isotropic photon''.
The concept of isotropic photon was used in the 
literature~\cite{Carra-95,Veenendaal-96-fluo,%
Veenendaal-96,Veenendaal-98,Carra-00,Ament-11}.
An isotropic incoming photon would correspond to 
$b=0$ in Eq.~\eqref{In} and an isotropic outgoing photon to $c=0$
in Eq.~\eqref{Out}.
At first, it was mistakenly stated that an 
isotropic photon is an 
unpolarized photon~\cite{Veenendaal-96}. As shown in Appendix~\ref{average},
this is not the case and an unpolarized incident photon
(for example) is the sum of a contribution $c=0$
and a contribution $c=2$. 
Indeed, the photon polarization vector $\epsilon$
is always perpendicular to the photon direction
$\bfk$. Thus, an unpolarized photon is not isotropic
in all space, it is only isotropic in the plane
perpendicular to $\bfk$. For example, angle-dependent
x-ray absorption can be carried out with unpolarized
beams \cite{BrouderAXAS}. This anisotropy of
unpolarized light is the origin of the $b=2$
contribution to the scattered beam.
The multipole (\textit{i.e.} $b=0$ and $b=2$) nature of unpolarized
light was clarified by Veenendaal and Benoist \cite{Veenendaal-98}. 
The identification of isotropic and unpolarized photons
was corrected in Ref.~\cite{Veenendaal-96}.

In the next sections,  RIXS spectra are
discussed for electric dipole emission
(i.e. $(\ell'_1,\ell'_2)=(0,0)$), and either
electric dipole excitation
(i.e. $(\ell_1,\ell_2)=(0,0)$)
or electric quadrupole excitation
(i.e. $(\ell_1,\ell_2)=(1,1)$).
Mixed excitations 
(i.e. $(\ell_1,\ell_2)=(0,1)$ and $(1,0)$)
are deferred to a forthcoming paper.

\subsection{Electric dipole excitation, electric dipole emission}
Let an electric dipole transition in the 
absorption be followed by an electric dipole transition in the emission: 
$\ell_1=0$, $\ell_2=0$, $\ell_1'=0$ and $\ell_2'=0$. Thus: 
\begin{eqnarray*}
0 \leq g_1 \leq 2,\quad
0 \leq g_2 \leq 2,\quad
0 \leq a \leq 4,\quad
0 \leq b \leq 2,\quad
0 \leq c \leq 2. 
\end{eqnarray*}
Since $\ell_1=0$, $\ell_2=0$, then $v=0$. This implies $c=u$. 
Since $\ell_1'=0$ and $\ell_2'=0$, then $v'=0$ and $b=u'$. 
These conditions allow the calculation of the values of all 
$9j$ symbols needed (see Appendix \ref{9je1}).  

The scattering cross section simplifies to:
\begin{eqnarray}
 \sigmarixs^{E1E1}
= \displaystyle\sum_{g_1,g_2}
  \displaystyle\sum_{a,b,c}(-1)^{a-g_2}
  \Pi_{g_1,g_2,b,c} \ninej{1~}{1~}{g_1}{1~}{1~}{g_2}{b}{c}{a}
\Big \{ \{ \epsilon^*_s \otimes \epsilon_s  \}^{(b)} \otimes \{ \epsilon \otimes \epsilon^*  \}^{(c)} \Big \}^{(a)} 
\cdot S^{g_1g_2 a}_{L_0}, 
\label{RIXSE1E1}
\end{eqnarray}
where $L_0=(0,0,0,0)$.

\subsection{Electric dipole excitation and dipole emission, isotropic sample}
A further simplification arises when the sample is isotropic. This corresponds to the case
of a liquid or a powder sample, with no prefered orientation and 
no remanent magnetization.
The ensuing isotropy implies that $a=0$ and
\begin{eqnarray*}
   \ninej{1~}{1~}{g_1}{1~}{1~}{g_2}{b}{c}{0} &=&
  \frac{\delta_{g_1 g_2}\delta_{bc} (-1)^{g_1+b}}{\sqrt{(2g_1+1)(2b+1)}}
   \sixj{1}{1}{g_1}{1}{1}{b}.
\end{eqnarray*}
Therefore,
\begin{eqnarray}
\sigmarixs^{E1E1}
&=& \displaystyle\sum_{g,b}
  (-1)^{b} \Pi_{g,b} 
   \sixj{1}{1}{g}{1}{1}{b} 
  \gamma_{L_0U}^{bb0}
    S^{gg0}_{L_0}.
\label{RIXSE1E1powder}
\end{eqnarray}
Note that the variables $g$ can take the values 0, 1 and 2, so that
only three fundamental spectra are needed to generate all the
spectra that can be measured on a powder.
The variable $b$ describing the incident and scattered beams can
take the value 0, 1 and 2.

\begin{itemize}
\item{For $b=0$:
\begin{eqnarray*}
\sigmarixs^{E1E1}(b=0)
&=& \sum_{g=0}^2 (-1)^g
  \frac{\sqrt{2g+1}}{9}
  S^{gg0}_{L_0}.
\end{eqnarray*}}
\item{For $b=1$:}
\begin{eqnarray*}
\{\epsilon\otimes\epsilon^*\}^{(1)} &=&
  \frac{i}{\sqrt{2}} \epsilon\times\epsilon^*
=-\frac{P_c}{\sqrt{2}} \bfk,
\end{eqnarray*}
where $P_c$ is the rate of circular polarization. Note that $P_c$
is positive for a right circular polarization in the traditional
sense (\textit{i.e.} for a negative helicity). 
$P_c=0$ if the polarization is linear. 
Similarly, $P_{c,s}$ is defined by
$P_{c,s}=i \epsilon^*_s\times\epsilon_s$ and
\begin{eqnarray*}
 \gamma_{L_0U}^{bb0} &=& \frac{1}{2\sqrt{3}}P_cP_{c,s} \bfk_s. \bfk
= \frac{1}{2\sqrt{3}}(|\epsilon\cdot\epsilon_s^*|^2
   -|\epsilon\cdot\epsilon_s|^2).
\end{eqnarray*}

Thus \begin{eqnarray}
\sigmarixs^{E1E1}(b=1)
&=& \sum_{g=0}^2 
 - \frac{\sqrt{2g+1}}{2} \sixj{1}{1}{g}{1}{1}{1} 
   (|\epsilon\cdot\epsilon_s^*|^2
   -|\epsilon\cdot\epsilon_s|^2) S^{gg0}_{L_0}.
\end{eqnarray}

This result is interesting: for an unoriented
sample, in particular for a powder sample without permanent
magnetization direction and in the absence of magnetic dipole transitions,
some circular dichroism can be observed
when measuring the circular polarization of the 
scattered beam, for example using the polarization analysis device described in Ref.~\cite{Ishii-2011}. However, 
the incident and scattered wavevectors must not be
perpendicular. This remark shows the power of the
geometric (coordinateless) approach. 
Note that $\epsilon\cdot\epsilon_s^*$
and $\epsilon\cdot\epsilon_s$ are 
expected to be involved in this formula
because they are the only non-trivial
scalars that can be built from $\epsilon$, $\epsilon_s$
and their conjugates. 

\item{For $b=2$:}
\begin{eqnarray*}
\sigmarixs^{E1E1}(b=2)
=   \sum_{g=0}^2 
  \frac{4\sqrt{2g+1}}{(2-g)!(3+g)!} 
     (\frac{1}{2}|\epsilon^* \cdot \epsilon_s|^2+
  \frac{1}{2}|\epsilon \cdot \epsilon_s|^2-\frac{1}{3})
  S^{gg0}_{L_0}.
\end{eqnarray*}

\end{itemize}
The final expression of the Kramers-Heisenberg cross-section for an isotropic sample, electric dipole absorption and dipole emission is:

\begin{eqnarray}
\sigmarixs^{E1E1}  &=&
\sum_{g=0}^2 
  \Big ((-1)^g
  \frac{\sqrt{2g+1}}{9} - \frac{\sqrt{2g+1}}{2} \sixj{1}{1}{g}{1}{1}{1} 
    (|\epsilon\cdot\epsilon_s^*|^2
   -|\epsilon\cdot\epsilon_s|^2) 
\nonumber\\&&
  +\frac{4\sqrt{2g+1}}{(2-g)!(3+g)!} 
    (\frac{1}{2}|\epsilon^* \cdot \epsilon_s|^2+\frac{1}{2}
  |\epsilon \cdot \epsilon_s|^2-\frac{1}{3}) \Big ) S^{gg0}_{L_0}.
  \label{e1e1}
\end{eqnarray}

If the polarization of the scattered beam is not detected, then the 
term $b=2$ is calculated using the relation:
$\langle \{\epsilon_s\otimes\epsilon_s^*\}^{(2)} \rangle 
= - \bfk_s^{(2)}/2$ (see Appendix \ref{average}) and

\begin{eqnarray}
   \Big\langle \gamma_{L_0U}^{bb0}
\Big\rangle
&=& -\frac{1}{2\sqrt{5}}\big( |\bfk_s \cdot \epsilon|^2-\frac{1}{3}\big).
\end{eqnarray}

In this case, 

\begin{eqnarray}
\Big\langle\sigmarixs^{E1E1} \Big\rangle\
= \sum_{g=0}^2 
  \Big ((-1)^g
  \frac{\sqrt{2g+1}}{9}  
 -\frac{2\sqrt{2g+1}}{(2-g)!(3+g)!} (|\bfk_s \cdot \epsilon|^2-\frac{1}{3}) \Big )
 S^{gg0}_{L_0}.
  \label{e1e1av}
\end{eqnarray}

\subsection{Electric quadrupole excitation, electric dipole emission}

This section considers the case of electric quadrupole transitions in the absorption followed by electric dipole transitions in the emission: $\ell_1=1$, $\ell_2=1$, $\ell_1'=0$ and $\ell_2'=0$. Thus: 

\begin{eqnarray*}
 &&1 \leq g_1 \leq 3, \quad
1 \leq g_2 \leq 3,\quad
0 \leq a \leq 6,\quad
0 \leq b \leq 2,\quad
0 \leq c \leq 4. 
\end {eqnarray*}
Since $\ell_1=1$ and $\ell_2=1$, $ 0 \leq v \leq 2$. Additionally, $v \neq 1$ since $\{ \bfk \otimes \bfk \} ^{(1)}= \frac{i}{\sqrt{2}}\bfk \times \bfk =0$. Thus, $v=0$ or 2. Since $\ell_1'=0$ and $\ell_2'=0$, $v'=0$ and  $b=u'$.  The values of all $9j$ symbols needed are given in Appendix~\ref{9je2}.  \\

Thus,
\begin{eqnarray}
\sigmarixs^{E2E1} = \sum_{g_1,g_2}
\sum_{a,b,c,u,v} \frac{5}{4} (-1)^{a+1-g_2} \Pi_{g_1,g_2,b,c,u,v} 
\ninej{1}{1}{2}{1}{1}{2}{u}{v}{c} \ninej{1}{2}{g_1}{1}{2}{g_2}{b}{c}{a}
\gamma^{bca}_{UL_1} S^{g_1g_2a}_{L_1},
  \label{scattfinal1}
\end{eqnarray}
where $L_1=(1,1,0,0)$.

\subsection{Electric quadrupole excitation, electric dipole emission, isotropic sample}

Isotropy implies that $a=0$. This further implies that $g_1=g_2$ and $b=c=u'$. 

\begin{eqnarray}
\sigmarixs^{E2E1} \nonumber = \sum_{g}
\sum_{b,u,v} \frac{5}{4} (-1)^{1-g} (2g+1)(2b+1)\Pi_{u,v} 
\ninej{1}{1}{2}{1}{1}{2}{u}{v}{b}
\ninej{1}{2}{g}{1}{2}{g}{b}{b}{0}
\gamma^{bb0}_{UL_1} S^{gg0}_{L_1},
  \label{scattfinal2}
\end{eqnarray}
with $U=(u,v,b,0)$.

As in the case of electric dipole absorption and emission, 
only three fundamental spectra are needed to describe all the experimental
spectra of a powder.  The tensors $\gamma^{bb0}_{UL_1}$,
are evaluated for all possible values of ($u,b,v$) (see Appendix~\ref{e2e2p}). This leads to the final expression of the Kramers-Heisenberg cross-section for an isotropic sample, in the case of quadrupole absorption with dipole emission:

\begin{eqnarray}
\sigmarixs^{E2E1}
&=& \sum_{g=1}^3 
  \Big ((-1)^g
  \frac{\sqrt{2g+1}}{120} + \frac{1}{16\sqrt{5}} \sqrt{2g+1} 
   \sixj{1}{2}{g}{2}{1}{1}  \times (|\epsilon\cdot\epsilon_s^*|^2
   -|\epsilon\cdot\epsilon_s|^2)  
  \nonumber\\  && 
+\frac{\sqrt{3}}{4\sqrt{7}}\sqrt{2g+1}\sixj{1}{2}{g}{2}{1}{2} \times
(\frac{1}{3}-\frac{1}{4}|\epsilon^* \cdot
\epsilon_s|^2-\frac{1}{4}|\epsilon \cdot
\epsilon_s|^2-\frac{1}{2}|\bfk\cdot\epsilon_s^*|^2) \Big )
S^{gg0}_{L_1},
\label{e2e1}
\end{eqnarray}
where: $\sixj{1}{2}{g}{2}{1}{2}=\frac{1}{5\sqrt{21}}$ for $g=3$, $\frac{\sqrt{7}}{10\sqrt{3}}$ for $g=2$, $\frac{\sqrt{7}}{10\sqrt{3}}$ for $g=1$,
$\sixj{1}{2}{g}{2}{1}{1}=\frac{1}{3\sqrt{5}}$ for $g=3$, $\frac{1}{6\sqrt{5}}$ for $g=2$, and $-\frac{1}{2\sqrt{5}}$ for $g=1$.

If the polarization of the scattered beam is not detected 
the polarization average yields:
\begin{eqnarray}
\Big\langle\sigmarixs^{E2E1} \Big\rangle\ 
 = \sum_{g=1}^3 
  \Big ((-1)^g
  \frac{\sqrt{2g+1}}{120} +\frac{\sqrt{3}}{4\sqrt{7}}\sqrt{2g+1}
   \sixj{1}{2}{g}{2}{1}{2}\times (-\frac{1}{6}
+\frac{1}{4}|\bfk \cdot \bfk_s|^2+\frac{1}{4}|\epsilon \cdot \bfk_s|^2) \Big )
S^{gg0}_{L_1}.
  \label{e2e1av}
 \end{eqnarray}

\section{Conclusion}
This paper presents a detailed derivation of the resonant inelastic scattering 
cross section in geometric spherical tensor form. Starting from the 
Kramers-Heisenberg equation and using angular-momentum coupling techniques, 
this expression is derived without any assumption on the 
nature of the states involved in the scattering process. 
The use of spherical tensors allows to drastically reduce the number of 
fundamental spectra that shall be measured in order to extract the full 
information from a sample. For example, for electric dipole absorption 
followed by electric dipole emission transitions, 19 fundamental spectra
(versus 81 for a general fourth-rank Cartesian tensor) are required,
that reduce to three for a powder.

Angular momentum techniques are used to recouple each photon 
(incident and scattered) with itself. This step is required 
to discuss the common case where the polarization of the
scattered beam is not measured.
 
Some special cases were studied to illustrate the ability of this expression 
to describe global properties of the sample. In the case of isotropic samples, 
which are most often measured experimentally, the electric dipole 
absorption-electric dipole emission and electric quadrupole absorption-electric 
dipole emission cross-sections are each expressed as a combination of only three fundamental spectra. 
This method predicts that 
circular dichroism may be observed on isotropic samples provided that 
the circular polarization of the scattered beam is detected.

\section*{Acknowledgements}
We are very grateful to Myrtille Hunault for her thorough reading
of the manuscript.

\section{Appendix}

\subsection{Derivation of Kramers-Heisenberg formula using spherical tensors}
\label{KHsphe}

\subsubsection{Expression of $\epsilon\cdot\bfr (\bfk\cdot\bfr)^{\ell}$\\}
\label{exp}

First it is shown that $\epsilon\cdot\bfr (\bfk\cdot\bfr)^{\ell}$ can be expressed 
as $g_{\ell}\Big\{\{\epsilon \otimes \bfk^{\ell}\}^{({\ell}+1)} \otimes 
\bfr^{({\ell}+1)}\Big\}^{(0)}$ where $g_{\ell}$ is a constant.

For $\ell=0$, according to Ref. \cite{Brouder} (Eq. 13):
\begin{eqnarray}
\epsilon\cdot\bfr
= - \sqrt{3}(-\frac{1}{\sqrt{3}}\epsilon\cdot\bfr)=- \sqrt{3}\{\epsilon\otimes\bfr\}^0
=- \sqrt{3}\Big\{\{\epsilon \otimes \bfk^{(0)}\}^{(1)} \otimes \bfr^{(1)}\Big\}^{(0)}.
\end{eqnarray} 

For $\ell=1$, according to Ref. \cite{Brouder}[p.~5]:
\begin{eqnarray}
\epsilon\cdot\bfr \bfk\cdot\bfr
 = \sqrt{5}\{\{\epsilon\otimes\bfk\}^{(2)} \otimes \{\bfr\otimes \bfr\}^{(2)}\}^{(0)} 
= \sqrt{5}\Big\{\{\epsilon\otimes\bfk\}^{(2)} \otimes
\bfr^{(2)}\Big\}^{(0)}.
\end{eqnarray} 

Thus, $\epsilon\cdot\bfr (\bfk\cdot\bfr)^{\ell}=g_{\ell}\Big\{\{\epsilon \otimes \bfk^{{\ell}}\}^{({\ell}+1)} \otimes \bfr^{({\ell}+1)}\Big\}^{(0)}$, with $\bfr^{(1)}=\bfr$, $g_0=-\sqrt{3}$, $\bfr^{(2)}=\{\bfr\otimes \bfr\}^{(2)}$ and $g_1=\sqrt{5}$. 

Defining $ E_{IFN}=\frac{(E_I-E_N)(E_N-E_F)}{E_I-E_N+\hbar\omega+i\gamma}$, 
Equation\eqref{scatcc3} becomes: 
\begin{eqnarray}
\nonumber
 \sigmarixs = C \sum_F
\Big|
\sum_N E_{IFN} \sum_{{\ell},{\ell'}=0}^{1}f_{\ell}f_{\ell'}^*g_{\ell}g_{\ell'}
 \langle N|\Big\{\{\epsilon \otimes \bfk^{{\ell}}\}^{({\ell}+1)} \otimes \bfr^{({\ell}+1)}\Big\}^{(0)}|I\rangle  \\
 \langle F|\Big\{\{\epsilon_s^* \otimes \bfk_s^{{\ell'}}\}^{({\ell'}+1)} \otimes \bfr^{({\ell'}+1)}\Big\}^{(0)}|N\rangle
   \Big|^2 \delta_{E},
  \end{eqnarray}
    and 
 \begin{eqnarray}
 \nonumber
 \sigmarixs = C \sum_F
\Big|
\sum_N E_{IFN} \sum_{{\ell},{\ell'}=0}^{1}h_{\ell}h_{\ell'}^*
 \Big\{\{\epsilon \otimes \bfk^{{\ell}}\}^{({\ell}+1)} \otimes \langle N|\bfr^{({\ell}+1)}|I\rangle \Big\}^{(0)} \\
 \Big\{\{\epsilon_s^* \otimes \bfk_s^{{\ell'}}\}^{({\ell'}+1)} \otimes \langle F|\bfr^{({\ell'}+1)}|N\rangle \Big\}^{(0)}
   \Big|^2 \delta_{E},
   \label{scatcc4}
\end{eqnarray}
with $h_{\ell}=f_{\ell}g_{\ell}$.
If one now defines $\bfr_{NI}^{(\ell+1)}= \langle N|\bfr^{({\ell}+1)}|I\rangle$ and 
$\bfr_{FN}^{(\ell'+1)}=\langle F|\bfr^{({\ell'}+1)}|N\rangle$,
one obtains: 
\begin{eqnarray}
 \sigmarixs =C \sum_F
\Big|
\sum_N E_{IFN} \sum_{{\ell},{\ell'}=0}^{1}h_{\ell}h_{\ell'}^*
 \Big\{\{\epsilon \otimes \bfk^{{\ell}}\}^{({\ell}+1)} \otimes \bfr_{NI}^{(\ell+1)} \Big\}^{(0)} 
 \Big\{\{\epsilon_s^* \otimes \bfk_s^{{\ell'}}\}^{({\ell'}+1)} \otimes \bfr_{FN}^{(\ell'+1)} \Big\}^{(0)}
   \Big|^2 \delta_{E}.
\label{scatcc5}
\end{eqnarray}

\subsubsection{Recoupling spherical tensors\\}
\label{couple}
Now using the recoupling identity (Ref.\cite{Brouder}, Eq.~14), which is:
\begin{eqnarray}
\{P^{(a)} \otimes Q^{(a)}\}^{(0)} \{R^{(d)} \otimes S^{(d)}\}^{(0)}
= \sum_{g=|a-d|}^{a+d}(-1)^g\frac{ \{P^{(a)} \otimes R^{(d)} \}^{(g)}\cdot \{Q^{(a)} \otimes S^{(d)} \}^{(g)}}{\sqrt{(2a+1)(2d+1)}}, 
\label{coupl1}
\end{eqnarray}
where the scalar product $\cdot$ of two spherical tensors $P^{(g)}$ and $Q^{(g)}$ is defined by:
$P^{(g)}\cdot Q^{(g)}=\sum_{{\gamma}=-g}^{g} (-1)^\gamma P^{(g)}_\gamma Q^{(g)}_{-\gamma}$,
Equation\eqref{scatcc5} becomes:
\begin{eqnarray*}
\sigmarixs =C \sum_F
\Big|
\sum_N \sum_{g,{\ell},{\ell'}}E_{IFN} \frac{(-1)^g h_{\ell}h_{\ell'}^*}{\sqrt{(2\ell+3)(2\ell'+3)}} 
\Big\{ \{\epsilon_s^* \otimes \bfk_s^{\ell'} \}^{(\ell'+1)} \otimes \{\epsilon \otimes \bfk^{\ell} \}^{(\ell+1)}\Big\}^{(g)} 
\cdot \Big\{\bfr_{FN}^{(\ell'+1)} \otimes  \bfr_{NI}^{(\ell+1)}\Big\}^{(g)}\Big|^2 \delta_{E},
\end{eqnarray*}
where $g$ runs from $|\ell-\ell'|$ to $(\ell+\ell'+2)$, $\ell$ and $\ell'$ run from 0 to 1. This is Equation\eqref{eq4}.

\subsubsection{Expansion of the square in Equation\eqref{sc6}}
\label{square1}
Now the modulus of the amplitude inside the sum over the final states in
Equation\eqref{sc6} is squared. When expanding the square, Equation\eqref{sc6} becomes:
\begin{eqnarray}
\nonumber\sigmarixs =  C \sum_F 
\sum_{g_1,{\ell}_1,{\ell'}_1,
g_2,{\ell}_2,{\ell'}_2} h_{\ell_1}h_{\ell'_1}^*h_{\ell_2}^*h_{\ell'_2}
\frac{(-1)^{g_1+\ell_2+\ell'_2}}{\sqrt{(2\ell_1+3)(2\ell'_1+3)(2\ell_2+3)(2\ell'_2+3)}} \\ \nonumber
 \Big\{ \{\epsilon_s^* \otimes 
\bfk_s^{\ell'_1} \}^{(\ell'_1+1)} \otimes \{\epsilon \otimes 
\bfk^{\ell_1}\}^{(\ell_1+1)}\Big\}^{(g_1)} \cdot A_{FI}^{(g_1)}(\ell_1, \ell'_1) 
\nonumber\\
 \Big\{ \{\epsilon_s \otimes \bfk_s^{\ell'_2} \}^{(\ell'_2+1)}
 \otimes \{\epsilon^* \otimes \bfk^{\ell_2} \}^{(\ell_2+1)}\Big\}^{(g_2)}  \cdot
  A_{FI}^{(g_2)}(\ell_2, \ell'_2)^* \delta_{E}, 
 \label{scatcc7}
\end{eqnarray}
where $g_i$ runs from $|\ell_i-\ell_i'|$ to $(\ell_i+\ell_i'+2)$, with $\ell_i$ and $\ell_i'$ equal to 0 or 1. In Appendix \ref{complex}, it is shown that $A_{FI}^{(g_2)}(\ell_2, \ell'_2)^*= \barA_{IF}^{(g_2)}(\ell_2, \ell'_2)$.

Therefore,
\begin{eqnarray}
\sigmarixs \nonumber = C\sum_F
\sum_{g_1,{\ell}_1,{\ell'}_1,
g_2,{\ell}_2,{\ell'}_2}h_{\ell_1}h_{\ell'_1}^*h_{\ell_2}^*h_{\ell'_2} \frac{(-1)^{g_1+\ell_2+\ell'_2} 
}{\sqrt{(2\ell_1+3)(2\ell'_1+3)(2\ell_2+3)(2\ell'_2+3)}} \\ \nonumber
 \Big\{ \{\epsilon_s^* \otimes 
\bfk_s^{\ell'_1} \}^{(\ell'_1+1)} \otimes \{\epsilon \otimes 
\bfk^{\ell_1}\}^{(\ell_1+1)}\Big\}^{(g_1)} \cdot 
A_{FI}^{(g_1)}(\ell_1, \ell'_1) \\
  \nonumber
 \Big\{ \{\epsilon_s \otimes \bfk_s^{\ell'_2} \}^{(\ell'_2+1)}
 \otimes \{\epsilon^* \otimes \bfk^{\ell_2} \}^{(\ell_2+1)}\Big\}^{(g_2)} \cdot
  \barA_{IF}^{(g_2)}(\ell_2, \ell'_2) \delta_{E}.
 \label{scatcc7b}
\end{eqnarray}  
  
If one defines
\begin{eqnarray}
X_1^{(g_1)}= \Big \{ \{\epsilon_s^* \otimes \bfk_s^{\ell'_1} \}^{(\ell'_1+1)} \otimes \{\epsilon \otimes \bfk^{\ell_1}\}^{(\ell_1+1)} \Big \}^{(g_1)},\\
X_2^{(g_2)}= \Big \{\{\epsilon_s \otimes \bfk_s^{\ell'_2} \}^{(\ell'_2+1)} \otimes \{\epsilon^* \otimes \bfk^{\ell_2} \}^{(\ell_2+1)} \Big \}^{(g_2)},\\
\Pi_{\ell_1+1,\ell_2+1,\ell'_1+1,\ell'_2+1}  =\sqrt{(2\ell_1+3)(2\ell'_1+3)(2\ell_2+3)(2\ell'_2+3)},
\end{eqnarray}

one obtains
\begin{eqnarray}
\sigmarixs
= C \sum_F
\displaystyle\sum_{g_1,{\ell}_1,{\ell'}_1, 
  g_2,{\ell}_2,{\ell'}_2} \frac{(-1)^{g_1+\ell_2+\ell'_2} 
  h_{\ell_1}h_{\ell'_1}^*h_{\ell_2}^*h_{\ell'_2}}
  {\Pi_{\ell_1+1,\ell_2+1,\ell'_1+1,\ell'_2+1}}  X_1 ^{(g_1)} \cdot 
  A_{FI}^{(g_1)}(\ell_1, \ell'_1)\,\,  X_2 ^{(g_2)} \cdot 
  \barA_{IF}^{(g_2)}(\ell_2, \ell'_2) \delta_{E}.
\label{scatcc8}
\end{eqnarray}

The identity $P^{(a)} \cdot Q^{(a)} = (-1)^a \sqrt{2a+1}~\{ P^{(a)} 
\otimes Q^{(a)} \}^{(0)}$~\cite[pp.~64-65]{Varsha}) becomes:
\begin{eqnarray*}
 X_1 ^{(g_1)} \cdot A_{FI}^{(g_1)}(\ell_1, \ell'_1) & = & 
(-1)^{g_1}\sqrt{2g_1+1}~\{X_1^{(g_1)} \otimes A_{FI}^{(g_1)}(\ell_1, \ell'_1) \}^{(0)},
\\
  X_2 ^{(g_2)} \cdot \barA_{IF}^{(g_2)}(\ell_2, \ell'_2) &  = &
(-1)^{g_2}\sqrt{2g_2+1}~\{X_2^{(g_2)} \otimes 
\barA_{IF}^{(g_2)}(\ell_2, \ell'_2) \}^{(0)},
\end{eqnarray*}
and then:
\begin{eqnarray*}
\sigmarixs = C \sum_F
\sum_{g_1,{\ell}_1,{\ell'}_1,
g_2,{\ell}_2,{\ell'}_2} \frac{(-1)^{\ell_2+\ell'_2-g_2}
h_{\ell_1}h_{\ell'_1}^*h_{\ell_2}^*h_{\ell'_2} }
{\Pi_{\ell_1+1,\ell_2+1,\ell'_1+1,\ell'_2+1}} \sqrt{(2g_1+1)(2g_2+1)}
\\
\{X_1 ^{(g_1)} \otimes A_{FI}^{(g_1)}(\ell_1, \ell'_1)\}^{(0)}
 \cdot
 \{X_2 ^{(g_2)} \otimes \barA_{IF}^{(g_2)}(\ell_2, \ell'_2)\}^{(0)} \delta_{E}. 
\end{eqnarray*}

Using Equation\eqref{coupl1}:
\begin{eqnarray}
\nonumber
\{X_1 ^{(g_1)} \otimes A_{FI}^{(g_1)}(\ell_1, \ell'_1)\}^{(0)} \{ X_2 ^{(g_2)}
\otimes \barA_{IF}^{(g_2)}(\ell_2, \ell'_2)\}^{(0)}  =
\sum_{a=|g_1-g_2|}^{g_1+g_2}\frac{(-1)^a}{\sqrt{(2g_1+1)(2g_2+1)}}  
\{ X_1^{(g_1)} \otimes X_2^{(g_2)} \}^{(a)} \\ \cdot 
\{ A_{FI}^{(g_1)}(\ell_1, \ell'_1) \otimes  
\barA_{IF}^{(g_2)}(\ell_2, \ell'_2)  \} ^{(a)}.
\label{scatcc10}
\end{eqnarray}

Then, 
\begin{eqnarray}
\sigmarixs  = C \sum_F
\displaystyle\sum_{g_1,\ell_1,\ell'_1, g_2,\ell_2,\ell'_2}
\frac{(-1)^{\ell_2+\ell'_2-g_2}h_{\ell_1}h_{\ell'_1}^*
 h_{\ell_2}^*h_{\ell'_2}}{\Pi_{\ell_1+1,\ell_2+1,\ell'_1+1,\ell'_2+1}} 
 \sum_a~ (-1)^a \{ X_1^{(g_1)} \otimes X_2^{(g_2)} \}^{(a)} \nonumber \\
  \cdot 
  \{ A_{FI}^{(g_1)}(\ell_1, \ell'_1) \otimes  
  \barA_{IF}^{(g_2)}(\ell_2, \ell'_2)  \} ^{(a)} \delta_{E},
\label{scatcc11} 
\end{eqnarray}
with $a$ running from $|g_1-g_2|$ to $g_1+g_2$.

In order to transform the tensor product: 
\begin{eqnarray}
\nonumber 
\{ X_1^{(g_1)} \otimes X_2^{(g_2)} \}^{(a)} =   \Big \{\big\{ \{\epsilon_s^* \otimes 
\bfk_s^{\ell'_1} \}^{(\ell'_1+1)} \otimes \{\epsilon \otimes \bfk^{\ell_1}\}^{(\ell_1+1)} 
\big \}^{(g_1)}   \otimes \big \{\{\epsilon_s \otimes 
\bfk_s^{\ell'_2} \}^{(\ell'_2+1)} \otimes \{\epsilon^* \otimes \bfk^{\ell_2} \}^{(\ell_2+1)} 
\big \}^{(g_2)} \Big \} ^{(a)},
\end{eqnarray}
one uses the following identity~\cite[p.~70]{Varsha}: 
\begin{eqnarray}
\nonumber
\Big \{ \{ P^{(a)} \otimes Q^{(b)} \}^{(c)} \otimes \{ R^{(d)} \otimes S^{(e)} \}^{(f)} 
\Big \}^{(k)}  = \sum_{g,h} \Pi_{c,f,g,h} \ninej{a}{b}{c}{d}{e}{f}{g}{h}{k} 
\Big \{ \{ P^{(a)} \otimes R^{(d)} \}^{(g)} \otimes 
\{ Q^{(b)} \otimes S^{(e)} \}^{(h)} \Big \}^{(k)}, 
\label{coupl2}
\end{eqnarray}
where $|a-d|\leq g \leq a+d$, $|b-e|\leq h \leq b+e$ and $|g-h|\leq k \leq g+h$. 
This recoupling transforms the coupling of the incident photon with the scattered 
photons into a coupling of each photon with itself.
The 9-$j$ symbol that appears in this formula
was overlooked in ref.~\cite{Carra-95}.
This yields: 
\begin{eqnarray*}
\{ X_1^{(g_1)} \otimes X_2^{(g_2)} \}^{(a)} = \sum_{b,c} \Pi_{g_1,g_2,b,c} \ninej{\ell_1'+1~}{\ell_1+1~}{g_1}{\ell_2'+1~}{\ell_2+1~}{g_2}{b}{c}{a}~\{ X\}^{(a)},
\end{eqnarray*}

with 
\begin{eqnarray*}
\{ X\}^{(a)} \nonumber = \Big \{\big \{ \{\epsilon_s^* \otimes \bfk_s^{\ell'_1} \}^{(\ell'_1+1)} \otimes \{\epsilon_s \otimes \bfk_s^{\ell'_2}\}^{(\ell'_2+1)} \big \}^{(b)} \otimes \big \{ \{\epsilon \otimes \bfk^{\ell_1} \}^{(\ell_1+1)} \otimes \{\epsilon^* \otimes \bfk^{\ell_2}\}^{(\ell_2+1)} \big \}^{(c)} \Big \}^{(a)}. 
\end{eqnarray*}\\

According to the triangular conditions, the $9j$-factor is zero if any of the following conditions
is not satisfied 
\begin{eqnarray*}
|\ell_1' -\ell_2'| \leq b \leq \ell_1' + \ell_2',\quad
|\ell_1 -\ell_2| \leq c \leq \ell_1 + \ell_2,\quad
|b-c| \leq a \leq b+c~~. 
\end{eqnarray*}

\subsubsection{Recoupling of $\{ X\}^{(a)}$ \\}
\label{square2}

In $\{ X\}^{(a)}$, the variables concerning the scattered beam ($\epsilon_s,\bfk_s$) 
are gathered in the first tensor product, while those concerning the incident beam 
are gathered in the second product.
However, to treat the case of a partially polarized incident or scattered beam,
it is required to couple the polarization vectors with themselves:
in
$\{ X\}^{(a)} = \Big \{ \{ X_{\mathrm{out}}\}^{(b)} \otimes \{ X_{\mathrm{in}}\}^{(c)} \Big
\}^{(a)}$ the polarization vectors in $\{ X_{\mathrm{out}}\}^{(b)}$ and $\{
X_{\mathrm{in}}\}^{(c)}$ are recoupled. 
According to Eq.\eqref{coupl2}, one obtains:
\begin{eqnarray*}
\{ X_{\mathrm{in}}\}^{(c)} &=& \big \{ \{\epsilon \otimes \bfk^{\ell_1} \}^{(\ell_1+1)} \otimes
\{\epsilon^* \otimes \bfk^{\ell_2}\}^{(\ell_2+1)} \big \}^{(c)} 
=\sum_{u,v}\Pi_{\ell_1+1,\ell_2+1,u,v}\ninej{1~}{\ell_1~}{\ell_1+1}{1~}{\ell_2~}{\ell_2+1}{u}{v}{c} 
\mathrm{In}^{(c)}_{UL},
\end{eqnarray*}
where $\mathrm{In}^{(c)}_{UL}= \Big \{ \{ \epsilon \otimes \epsilon^*  \}^{(u)} \otimes 
\{ \bfk^{\ell_1} \otimes \bfk^{\ell_2} \} ^{(v)}\Big\}^{(c)}$ and
\begin{eqnarray*}
\{ X_{\mathrm{out}}\}^{(b)} &=& \big \{ \{\epsilon_s^* \otimes 
\bfk_s^{\ell'_1} \}^{(\ell'_1+1)} \otimes \{\epsilon_s \otimes 
\bfk_s^{\ell'_2}\}^{(\ell'_2+1)} \big \}^{(b)} 
=\sum_{u',v'}\Pi_{\ell'_1+1,\ell'_2+1,u',v'}
\ninej{1~}{\ell'_1~}{\ell'_1+1}{1~}{\ell'_2~}{\ell'_2+1}{u'}{v'}{b} 
\mathrm{Out}^{(b)}_{UL},
\label{scatt10}
\end{eqnarray*}
where $\mathrm{Out}^{(b)}_{UL} = \Big \{ \{ \epsilon_s^* \otimes \epsilon_s  \}^{(u')} 
\otimes \{ \bfk_s^{\ell'_1} \otimes \bfk_s^{\ell'_2} \} ^{(v')}\Big\}^{(b)}$.
In these expressions the multi-indices $U$ and $L$ stand for
$U=(u,v,u',v')$ and $L=(\ell_1,\ell_2,\ell'_1,\ell'_2)$.
A similar recoupling of the polarization vector with itself was also
carried out by Veenendaal and Benoist~\cite{Veenendaal-98}.

The $9j$-factors vanish if any of the following triangular conditions is not fullfilled:
\begin{eqnarray*}
0 \leq u \leq  2,\quad
|\ell_1-\ell_2| \leq v \leq  \ell_1+\ell_2, \quad
0 \leq u' \leq 2,\quad
|\ell'_1-\ell'_2| \leq v' \leq \ell'_1+\ell'_2.
\end{eqnarray*}

\subsection{Complex conjugate}
\label{complex}

It is required to calculate the complex conjugate of 
$A_{FI}^{(g)}(\ell, \ell')$.
The position operator $x$ is Hermitian. Therefore
\begin{eqnarray*}
\langle N |x | I\rangle &=& 
\langle N |x^\dagger | I\rangle 
= \langle I |x | N\rangle^*,
\end{eqnarray*}
and 
$\langle N |x | I\rangle^*=\langle I |x | N\rangle$.
The same is true for $y$ and $z$.
For the corresponing spherical tensors, 
\begin{eqnarray*}
\langle N |r^{(1)}_1 | I\rangle^* =
-(1/\sqrt{2})
\big(\langle N |x | I\rangle +i
\langle N |y | I\rangle\big)^* 
=
-(1/\sqrt{2})
\big(\langle I |x | N\rangle -i
\langle I |y | N\rangle\big)
=- \langle I |r^{(1)}_{-1} | N\rangle^*.
\end{eqnarray*}
An analogous calculation for the other components of $r^{(1)}$
gives
$\langle N |r^{(1)}_\lambda | I\rangle^* = (-1)^\lambda
\langle I |r^{(1)}_{-\lambda} | N\rangle$.
For $\ell=2$, 
\begin{eqnarray*}
\langle N |r^{(2)}_\mu | I\rangle^* &=& 
\sum_{\lambda\lambda'} (1\lambda 1\lambda'|2\mu)
\langle N |r^{(1)}_{\lambda} r^{(1)}_{\lambda'} | I\rangle^* 
=
\sum_{\lambda\lambda'} (1\lambda 1\lambda'|2\mu)
\sum_K
\langle N |r^{(1)}_{\lambda}|K\rangle^*
\langle K| r^{(1)}_{\lambda'} | I\rangle^* 
\\&=&
\sum_{\lambda\lambda'} (1\lambda 1\lambda'|2\mu) (-1)^{\lambda+\lambda'}
\sum_K
\langle K |r^{(1)}_{-\lambda}|N\rangle
\langle I| r^{(1)}_{-\lambda'} | K\rangle.
\end{eqnarray*}
The Clebsch-Gordan coefficient $(1\lambda 1\lambda'|2\mu)$
implies that $\lambda+\lambda'=\mu$.
By replacing $\lambda$ and $\lambda'$ by $-\lambda'$ and $-\lambda$,
respectively, the following is obtained
\begin{eqnarray*}
\langle N |r^{(2)}_\mu | I\rangle^* &=&  (-1)^{\mu}
\sum_{\lambda\lambda'} (1\,{-\lambda'} 1\,{-\lambda}|2\mu) 
\sum_K
\langle I| r^{(1)}_{\lambda} | K\rangle
\langle K |r^{(1)}_{\lambda'}|N\rangle\\
&=&  (-1)^{\mu}
\sum_{\lambda\lambda'} (1\,{-\lambda'} 1\,{-\lambda}|2\mu) 
\langle I| r^{(1)}_{\lambda} r^{(1)}_{\lambda'}|N\rangle.
\end{eqnarray*}
Now, the symmetry
$(\ell_1 m_1 \ell_2 m_2|\ell_3 m_3)=(\ell_2 -m_2 \ell_1 -m_1|\ell_3
-m_3)$ is used to obtain
\begin{eqnarray*} \langle N |r^{(2)}_\mu | I\rangle^* &=& (-1)^{\mu}
\sum_{\lambda\lambda'} (1\lambda 1\lambda'|2-\mu)
\langle I| r^{(1)}_{\lambda} r^{(1)}_{\lambda'}|N\rangle
= (-1)^\mu 
\langle I |r^{(2)}_{-\mu} | N\rangle.
\end{eqnarray*}
A recursive use of this argument leads to
$\langle N |r^{(\ell)}_m | I\rangle^* = 
(-1)^m 
\langle I |r^{(\ell)}_{-m} | N\rangle$
for any $\ell$.

A similar calculation can be carried out for
\begin{eqnarray*}
X_\gamma &=& \{\bfr_{FN}^{(\ell')} \otimes \bfr_{NI}^{(\ell)}\}^{(g)}_\gamma
=\sum_{m'm} (\ell' m' \ell m|g\gamma)
\langle F|\bfr^{(\ell')}_{m'}|N\rangle  
\langle N|\bfr^{(\ell)}_{m}|I\rangle.
\end{eqnarray*}
The complex conjugate of $X_\gamma$ is
\begin{eqnarray*}
X_\gamma^* &=& 
\sum_{m'm} (\ell' m' \ell m|g\gamma)
\langle F|\bfr^{(\ell')}_{m'}|N\rangle^*
\langle N|\bfr^{(\ell)}_{m}|I\rangle^* \nonumber \\
&=&
\sum_{m'm} (\ell' m' \ell m|g\gamma) (-1)^{m+m'}
\langle N|\bfr^{(\ell')}_{-m'}|F\rangle
\langle I|\bfr^{(\ell)}_{-m}|N\rangle.
\end{eqnarray*}
The same reasoning as for the complex conjugate of
$\langle N |r^{(2)}_\mu | I\rangle$ gives 
\begin{eqnarray*}
\Big(\{\bfr_{FN}^{(\ell')} \otimes
\bfr_{NI}^{(\ell)}\}^{(g)}_\gamma\Big)^*
&=& (-1)^\gamma
\{\bfr_{IN}^{(\ell)} \otimes \bfr_{NF}^{(\ell')}\}^{(g)}_{-\gamma}.
\end{eqnarray*}
Finally,
\begin{eqnarray}
A_{FI,\gamma}^{(g)}(\ell, \ell')^* 
  = (-1)^\gamma
\sum_N \frac{(E_I-E_N)(E_N-E_F)}{E_I-E_N+\hbar\omega-i\gamma}
 \{\bfr_{IN}^{(\ell+1)} \otimes  
    \bfr_{NF}^{(\ell'+1)}\}^{(g)}_{-\gamma}.
\end{eqnarray}
Therefore, defining 
\begin{eqnarray}
\barA_{IF}^{(g)}(\ell, \ell') 
  &=&
\sum_N \frac{(E_I-E_N)(E_N-E_F)}{E_I-E_N+\hbar\omega-i\gamma}
 \{\bfr_{IN}^{(\ell+1)} \otimes  
    \bfr_{NF}^{(\ell'+1)}\}^{(g)},
\end{eqnarray}
the final result
$A_{FI,\gamma}^{(g)}(\ell, \ell')^* = (-1)^\gamma
\barA_{IF,-\gamma}^{(g)}(\ell, \ell')$ follows.

The relation
$\langle N |r^{(\ell)}_\lambda|I\rangle = (-1)^\lambda
\langle I |r^{(\ell)}_{-\lambda}|N\rangle^*$
is standard and it just remains to calculate
the complex conjugate of
$\Big\{ \{\epsilon_s^* \otimes \bfk_s^{\ell'} \}^{(\ell'+1)} 
\otimes \{\epsilon \otimes \bfk^{\ell}
\}^{(\ell+1)}\Big\}^{(g)}_\gamma$.
The polarization vectors $\epsilon$ and $\epsilon_s$ are complex.
For a complex vector $\bfz=\bfa+i\bfb$,
where $\bfa=(a_x,a_y,a_z)$ and $\bfb=(b_x,b_y,b_z)$ are real,
the complex conjugate of the spherical component $\bfz^{(1)}_\lambda$ 
for $\lambda=1$ is
\begin{eqnarray*}
\big(\bfz^{(1)}_1\big)^* &=& -\frac{1}{\sqrt{2}}
\big( a_x + i b_x +i (a_y+i b_y)\big)^*
=
 -\frac{1}{\sqrt{2}}
\big( a_x - i b_x -i a_y- b_y)\big)
\\&=&
 -\frac{1}{\sqrt{2}}
\big( a_x - i b_x -i (a_y-i b_y)\big)
=- (\bfz^*)^{(1)}_{-1}.
\end{eqnarray*}
The calculation of the other components gives
$\big(\bfz^{(1)}_\lambda\big)^* 
=(-)^\lambda (\bfz^*)^{(1)}_{-\lambda}$.
Therefore,
$(\epsilon_\lambda)^*=(-1)^\lambda (\epsilon^*)_{-\lambda}$
and
$\big((\epsilon_s^*)_\lambda\big)^*=(-1)^\lambda
(\epsilon_s)_{-\lambda}$.
The vector $\bfk$ is real and 
$(\bfk^{(\ell)}_m)^*=(-1)^m \bfk^{(\ell)}_{-m}$.
The same proof as for $A^{(g)}_{FI}(\ell,\ell')$ leads to
\begin{eqnarray*}
\Big(\{\epsilon \otimes \bfk^{\ell} \}^{(\ell+1)}_m\Big)^* &=& (-1)^m
\sum_{\lambda\mu} (1\,{-\lambda}\, \ell\,{-\mu} | \ell+1,m)
\epsilon^*_\lambda k^{(\ell)}_\mu.
\end{eqnarray*}
Since $1\lambda$ and $\ell\mu$ are not interchanged,
another symmetry relation must be used
\begin{eqnarray}
(\ell_1 m_1 \ell_2 m_2|\ell_3 m_3) &=& (-1)^{\ell_1+\ell_2-\ell_3}
(\ell_1\,{-m_1} \ell_2\, {-m_2}|\ell_3\, {-m_3}),
\label{symrel}
\end{eqnarray}
to obtain
\begin{eqnarray*}
\Big(\{\epsilon \otimes \bfk^{\ell} \}^{(\ell+1)}_m\Big)^* =
(-1)^m
\sum_{\lambda\mu} (1\lambda \ell\mu | \ell+1,{-m})
\epsilon^*_\lambda k^{(\ell)}_\mu =
(-1)^m
\{\epsilon^* \otimes \bfk^{\ell} \}^{(\ell+1)}_{-m}.
\end{eqnarray*}
Here, the symmetry relation does not bring any additional
sign because $\ell_3=\ell_1+\ell_2$.
Similarly,
$\Big(\{\epsilon_s^*~\otimes~\bfk_s^{\ell'} \}^{(\ell'+1)}_{m'}\Big)^*~=
(-1)^{m'} \{\epsilon_s \otimes \bfk_s^{\ell'} \}^{(\ell'+1)}_{-m'}$.

In the calculation of the complex conjugate of 
$\Big\{ \{\epsilon_s^* \otimes \bfk_s^{\ell'} \}^{(\ell'+1)} 
\otimes \{\epsilon \otimes \bfk^{\ell}
\}^{(\ell+1)}\Big\}^{(g)}_\gamma$, the
sign $(-1)^{\ell+\ell'-g}$ must be retained to obtain
\begin{eqnarray*}
\Big(\Big\{ \{\epsilon_s^* \otimes \bfk_s^{\ell'} \}^{(\ell'+1)} 
\otimes \{\epsilon \otimes \bfk^{\ell}
\}^{(\ell+1)}\Big\}^{(g)}_\gamma\Big)^* 
= (-1)^\gamma (-1)^{\ell+\ell'-g}
\Big\{ \{\epsilon_s \otimes \bfk_s^{\ell'} \}^{(\ell'+1)} 
\otimes \{\epsilon^* \otimes \bfk^{\ell}
\}^{(\ell+1)}\Big\}^{(g)}_{-\gamma}.
\end{eqnarray*}
Finally, the complex conjugate of
$X=\Big\{ \{\epsilon_s^* \otimes \bfk_s^{\ell'} \}^{(\ell'+1)} 
\otimes \{\epsilon \otimes \bfk^{\ell}
\}^{(\ell+1)}\Big\}^{(g)}\cdot A_{FI}^{(g)}(\ell,\ell')$ is
\begin{eqnarray*}
X^*
&=& (-1)^{\ell+\ell'-g}
\Big\{ \{\epsilon_s \otimes \bfk_s^{\ell'} \}^{(\ell'+1)} 
\otimes \{\epsilon^* \otimes \bfk^{\ell}
\}^{(\ell+1)}\Big\}^{(g)}\cdot\barA_{IF}^{(g)}(\ell,\ell').
\end{eqnarray*}

\subsection{Average over polarizations}
\label{average}
One of the most powerful features of the geometric expressions
is that they can be calculated in a specific coordinate
system adapted to a particular problem, and then 
be valid in any system.
This is illustrated by describing the average
of the coupling of $\epsilon$ and $\epsilon^*$:
\begin{eqnarray*}
\langle \{\epsilon\otimes\epsilon^*\}^{(0)} \rangle &=& - \frac{1}{\sqrt{3}},\\
\langle \{\epsilon\otimes\epsilon^*\}^{(1)} \rangle &=& 0,\\
\langle \{\epsilon\otimes\epsilon^*\}^{(2)} \rangle &=& -
\frac{\bfk^{(2)}}{\sqrt{6}}
= -\sqrt{\frac{2\pi}{15}} Y_2(\bfk).
\end{eqnarray*}
Taking a reference frame where $\bfk$ is along $Oz$, the 
linear polarization vector is
$\epsilon=(\cos\psi,\sin\psi,0)$ and the corresponding
spherical tensor components are
$\epsilon^{(1)}_{\pm 1}=\mp \ee^{\pm i\psi}/\sqrt{2}$,
$\epsilon^{(1)}_{0}=0$.
Therefore, it is easy to calculate
\begin{eqnarray*}
\{\epsilon\otimes\epsilon^*\}^{(2)}_{\pm 2} &=& \frac{\ee^{\pm
2i\psi}}{2},\\
\{\epsilon\otimes\epsilon^*\}^{(2)}_{\pm 1} &=& 0,\\
\{\epsilon\otimes\epsilon^*\}^{(2)}_{0} &=& -\frac{1}{\sqrt{6}}.
\end{eqnarray*}
The average over two perpendicular polarizations
($\psi$ and $\psi+\pi/2$) or the average over 
all $\psi$ gives
\begin{eqnarray*}
\langle\{\epsilon\otimes\epsilon^*\}^{(2)}_{\pm 2}\rangle &=& 0,\\
\langle\{\epsilon\otimes\epsilon^*\}^{(2)}_{\pm 1}\rangle &=& 0,\\
\langle\{\epsilon\otimes\epsilon^*\}^{(2)}_{0}\rangle &=& -\frac{1}{\sqrt{6}}.
\end{eqnarray*}
In that reference frame 
$\bfk_m^{(2)} = \delta_{m,0}\sqrt{2/3}$. Thus,
$\langle\{\epsilon\otimes\epsilon^*\}^{(2)}\rangle = -
\bfk^{(2)}/2$ and this relation is true in
any reference frame because it is a relation between two
tensors.

Considering elliptically-polarized x-rays, the most general 
polarization vector in a frame where $\bfk$
is along $Oz$ is
\begin{eqnarray*}
\epsilon &=& \left(\begin{array}{c}
  \cos\chi \cos\psi + i \sin\chi \sin\psi \\
  \cos\chi\sin\psi -i \sin\chi \cos\psi\\
  0
   \end{array}\right),
\end{eqnarray*}
for which the degree of circular polarization is
$\sin 2\chi$. In particular, $\chi=\pi/4$
for a fully circularly polarized x-ray.
Then,
\begin{eqnarray*}
\{\epsilon\otimes\epsilon^*\}^{(1)}_{\pm 1} &=& 0,\\
\{\epsilon\otimes\epsilon^*\}^{(1)}_{0} &=& -\frac{1}{\sqrt{2}}
\sin 2\chi,
\end{eqnarray*}
and
\begin{eqnarray*}
\{\epsilon\otimes\epsilon^*\}^{(2)}_{\pm 2} &=& \frac{1}{2}
  \cos 2\chi \ee^{\pm 2i\psi},\\
\{\epsilon\otimes\epsilon^*\}^{(2)}_{\pm 1} &=& 0,\\
\{\epsilon\otimes\epsilon^*\}^{(2)}_{0} &=& -\frac{1}{\sqrt{6}}.
\end{eqnarray*}
Therefore, 
by using left and right fully circularly polarized 
beams, the same average as with linear
polarization is obtained.

Similarly, it can be shown that
$\langle |\bfa\cdot\epsilon|^2\rangle =(|\bfa|^2-|\bfa\cdot\bfk|^2)/2$.

\subsection{Values of $9j$-factors and geometrical coefficients for
particular cases of Equation\eqref{sc6}}

\subsubsection{Electric dipole excitation, electric dipole emission}
\label{9je1}

\begin{itemize}
\item{The first $9j$-factor is:
   $\ninej{1~}{\ell_1~}{\ell_1+1}{1~}{\ell_2~}{\ell_2+1}{u}{v}{c}= 
   \ninej{1}{0}{1}{1}{0}{1}{c}{0}{c}= \frac{1}{3\sqrt{2c+1}}$.}
\item{The second $9j$-factor is:
   $\ninej{1~}{\ell'_1~}{\ell'_1+1}{1~}{\ell'_2~}{\ell'_2+1}{u'}{v'}{c'}= 
   \ninej{1}{0}{1}{1}{0}{1}{b}{0}{b}= \frac{1}{3\sqrt{2b+1}}$.}
\end{itemize}

Thus:
\begin{eqnarray}
\sigmarixs^{E1E1}
&=& \sum_{g_1,g_2}
  \displaystyle\sum_{a,b,c}(-1)^{a-g_2}
  \Pi_{g_1,g_2,b,c} \ninej{1~}{1~}{g_1}{1~}{1~}{g_2}{b}{c}{a}
  \gamma^{bca}_{UL_0}\cdot
  S^{g_1g_2a}_{L_0},
\label{RIXSE1E12}
\end{eqnarray}
where $L_0=(0,0,0,0)$ because $\ell_1=\ell_2=\ell'_1=\ell'_2=0$ for electric dipole
emission and absorption and $U=(b,0,c,0)$.
Additionnally, recall that
$\gamma^{bca}_{UL_0}=\{\mathrm{Out}_{UL}^{(b)}\otimes \mathrm{In}_{UL_0}^{(c)}\}^{(a)}$,
where
$\mathrm{Out}_{UL}^{(b)} = \{ \epsilon^*_s \otimes \epsilon_s  \}^{(b)}$ and
$\mathrm{In}_{UL_0}^{(c)}=\{ \epsilon \otimes \epsilon^*  \}^{(c)}$.
Further simplifications arise when the sample is a powder
(\textit{i.e.} $a=0$).

It can be shown that:
\begin{eqnarray*}
   \ninej{1~}{1~}{g_1}{1~}{1~}{g_2}{b}{c}{0} &=&
  \frac{\delta_{g_1 g_2}\delta_{bc} (-1)^{g_1+b}}{\sqrt{(2g_1+1)(2b+1)}}
   \sixj{1}{1}{g_1}{1}{1}{b}.
\end{eqnarray*}

\begin{itemize}
\item{For $b=0$:
\begin{eqnarray*}
   \sixj{1}{1}{g}{1}{1}{0} &=& \frac{(-1)^g}{3}, 
\end{eqnarray*}
with $g=0,1,2$, and $\gamma_{UL_0}^{000} = \frac{1}{3}$.

Thus,
\begin{eqnarray*}
\sigmarixs^{E1E1}(b=0)
&=&  \sum_{g=0}^2 (-1)^g
  \frac{\sqrt{2g+1}}{9} S^{gg0}_{L_0}.
\end{eqnarray*}}
\item{For $b=1$:}
\begin{eqnarray*}
   \sixj{1}{1}{1}{1}{1}{1} &=& \sixj{1}{1}{2}{1}{1}{1}=\frac{1}{6}, 
\end{eqnarray*}
\begin{eqnarray*}
   \sixj{1}{1}{0}{1}{1}{1} &=& -\frac{1}{3}.
   \end{eqnarray*}
\begin{eqnarray*}
\{\epsilon\otimes\epsilon^*\}^{(1)} &=&
  \frac{i}{\sqrt{2}} \epsilon\times\epsilon^*
=-\frac{P_c}{\sqrt{2}} \bfk,
\end{eqnarray*}
where $P_c$ is the rate of circular polarization. 

\begin{eqnarray*}
\gamma_{UL_0}^{110} &=& \frac{1}{2\sqrt{3}}P_cP_{c,s} \bfk_s\cdot \bfk
= \frac{1}{2\sqrt{3}}(|\epsilon\cdot\epsilon_s^*|^2
   -|\epsilon\cdot\epsilon_s|^2).
\end{eqnarray*}

Thus,
\begin{eqnarray*}
\sigmarixs^{E1E1}(b=1)
&=& \sum_{g=0}^2 
 - \frac{\sqrt{2g+1}}{2} \sixj{1}{1}{g}{1}{1}{1} (|\epsilon\cdot\epsilon_s^*|^2
   -|\epsilon\cdot\epsilon_s|^2)
   S^{gg0}_{L_0}.
\end{eqnarray*}

\item{For $b=2$:}
\begin{eqnarray*}
   \sixj{1}{1}{g}{1}{1}{2} &=& \frac{4}{(2-g)!(3+g)!}, 
\end{eqnarray*}
with $g=0,1,2$.

\begin{eqnarray*}
\gamma_{UL_0}^{220} &=& 
\frac{1}{\sqrt{5}}\big( \frac{1}{2}|\epsilon^* \cdot \epsilon_s|^2
  +\frac{1}{2}|\epsilon \cdot \epsilon_s|^2-\frac{1}{3}\big).
\end{eqnarray*}

Thus,
\begin{eqnarray*}
\sigmarixs^{E1E1}(b=2)
=   \sum_{g=0}^2 
  \frac{4\sqrt{2g+1}}{(2-g)!(3+g)!} (\frac{1}{2}|\epsilon^* \cdot \epsilon_s|^2
  +\frac{1}{2}|\epsilon \cdot \epsilon_s|^2-\frac{1}{3}) S^{gg0}_{L_0}.
\end{eqnarray*}
\end{itemize}

This leads to Equation\eqref{e1e1}.

If the polarization of the scattered beam is not detected, then the 
term $b=2$ is calculated using the relation:
$\langle \{\epsilon_s\otimes\epsilon_s^*\}^{(2)} \rangle 
= - \bfk_s^{(2)}/2$  and
 Eq.~\eqref{e1e1av} is obtained because
\begin{eqnarray}
   \Big\langle \gamma_{UL_0}^{220}
\Big\rangle
&=& -\frac{1}{2\sqrt{5}}\big( |\bfk_s \cdot \epsilon|^2-\frac{1}{3}\big).
\end{eqnarray}

\subsubsection{Electric quadrupole excitation, electric dipole emission}
\label{9je2}
In the case of electric quadrupole transitions in the absorption followed by electric dipole transitions in the emission, $\ell_1=1$, $\ell_2=1$, $\ell_1'=0$ and $\ell_2'=0$. Thus, 
\begin{eqnarray*}
1 \leq g_1 \leq 3, \quad
1 \leq g_2 \leq 3,\quad
0 \leq a \leq 6,\quad
0 \leq b \leq 2,\quad
0 \leq c \leq 4. 
\end {eqnarray*}
Since $\ell_1=1$ and $\ell_2=1$, $ 0 \leq v \leq 2$. Additionally, $v \neq 1$ since $\{ \bfk^{1} \otimes \bfk^{1} \} ^{(1)}= \frac{i}{\sqrt{2}}\bfk \times \bfk =0$. Thus, $v=0$ or 2. \\
Since $\ell_1'=0$ and $\ell_2'=0$, $v'=0$ and  $b=u'$.   \\

Therefore:
\begin{itemize}
\item{The first $9j$-factor is:$\ninej{1~}{\ell_1~}{\ell_1+1}{1~}{\ell_2~}{\ell_2+1}{u}{v}{c}= \ninej{1}{1}{2}{1}{1}{2}{u}{v}{c}$.
It is zero if $u+v+c$ is odd.}
\item{The second $9j$-factor is: $\ninej{1~}{\ell'_1~}{\ell'_1+1}{1~}{\ell'_2~}{\ell'_2+1}{u'}{v'}{b}= \ninej{1}{0}{1}{1}{0}{1}{b}{0}{b}=\frac{1}{3\sqrt{2b+1}}$.}
\item{The third $9j$-factor is:
$\ninej{\ell'_1+1~}{\ell_1+1~}{g_1}{\ell'_2+1~}{\ell_2+1~}{g_2}{b}{c}{a}= \ninej{1}{2}{g_1}{1}{2}{g_2}{b}{c}{a}$.}
\end{itemize}

\subsubsection{Electric quadrupole excitation, electric dipole emission, powder sample}
\label{e2e2p}
Isotropy implies that $a=0$, $g_1=g_2$ and $b=c=u'$. 
\begin{eqnarray}
\nonumber
\sigmarixs^{E2E1}  = C \sum_F\displaystyle\sum_{g}
\sum_{b,u,v}(-1)^{1-g} h_{1}h_{0}^*
h_{1}^*h_{0} \Pi_{g,g,b,b,u,v,b} 
\small{\ninej{1}{1}{2}{1}{1}{2}{u}{v}{b}
\ninej{1}{0}{1}{1}{0}{1}{b}{0}{b}}
\ninej{1}{2}{g}{1}{2}{g}{b}{b}{0}
\gamma^{bb0}_{UL_1} S^{gg0}_{L_1},
  \label{scattfinal3}
\end{eqnarray}
with 
$L_1=(1,1,0,0)$ and  
$U=(u,v,u',v') = (u,v,b,0)$.
The angular term is
\begin{eqnarray*}
\gamma^{bb0}_{UL_1} &=& 
\Big\{ \big \{ \{ \epsilon \otimes \epsilon^*  \}^{(u)} \otimes 
    \{ \bfk \otimes \bfk \} ^{(v)}\big\}^{(b)} \otimes
\{ \epsilon_s^* \otimes \epsilon_s  \}^{(b)}\Big\}^{(0)}.
\label{sca22}
\end{eqnarray*}
More precisely,
\begin{itemize}
\item $b=0,u=0,v=0$: 
$\gamma^{bb0}_{UL_1} =-\frac{1}{3\sqrt{3}}$.
\item $b=0,u=2,v=2$: $\gamma^{bb0}_{UL_1} =\frac{1}{3\sqrt{15}}.$
\item $b=1,u=1,v=0$: $\gamma^{bb0}_{UL_1} =-\frac{1}{6}P_cP_{c,s}\bfk \cdot \bfk_s.$
\item $b=1,u=1,v=2$: $\gamma^{bb0}_{UL_1} =-\frac{1}{3\sqrt{5}}
    P_cP_{c,s}\bfk \cdot \bfk_s.$
\item $b=2,u=2,v=0$: $\gamma^{bb0}_{UL_1} =-\frac{1}{\sqrt{15}}
    \big( \frac{1}{2}|\epsilon^* \cdot \epsilon_s|^2
    +\frac{1}{2}|\epsilon \cdot \epsilon_s|^2-\frac{1}{3}\big).$
\item $b=2,u=0,v=2$: $\gamma^{bb0}_{UL_1} =-\frac{1}{\sqrt{15}}
   \big(|\bfk \cdot \epsilon_s|^2-\frac{1}{3}\big).$
\item $b=2,u=2,v=2$: $\gamma^{bb0}_{UL_1} =\frac{1}{3\sqrt{105}}
  (6|\bfk \cdot \epsilon_s|^2 - 4 + 3 |\epsilon \cdot \epsilon_s^*|^2
+ 3 |\epsilon \cdot \epsilon_s|^2).$
\end{itemize}

This leads to Equation\eqref{e2e1}.
Note that the term $b=2,u=2,v=2$ is obtained by using classical invariant theory~\cite{Weyl-Group,Lehman-89}: 

If the polarization of the scattered beam is not detected, 
then $\Big\langle \gamma^{110}_{UL_1} \big\rangle = 0$
and the polarization average yields:

\begin{eqnarray*}
\Big\langle \Big\{ \big \{ \{ \epsilon \otimes \epsilon^*  \}^{(2)} 
     \otimes \{ \bfk \otimes \bfk \} ^{(2)}\big\}^{(2)}\otimes
\{ \epsilon_s^* \otimes \epsilon_s  \}^{(2)}\Big\}^{(0)} \Big\rangle
=
\frac{2-3|\bfk \cdot \bfk_s|^2 - 3|\epsilon \cdot \bfk_s|^2}{3\sqrt{105}}.
\end{eqnarray*}

\begin{eqnarray*}
\Big\langle \Big\{ \big \{ \{ \epsilon \otimes \epsilon^*  \}^{(0)} 
     \otimes \{ \bfk \otimes \bfk \} ^{(2)}\big\}^{(2)}\otimes
\{ \epsilon_s^* \otimes \epsilon_s  \}^{(2)}\Big\}^{(0)} \Big\rangle
=
-\frac{1}{2\sqrt{15}}(\frac{1}{3} - |\bfk \cdot \bfk_s|^2).
\end{eqnarray*}

\begin{eqnarray*}
\Big\langle \Big\{ \big \{ \{ \epsilon \otimes \epsilon^*  \}^{(2)} 
     \otimes \{ \bfk \otimes \bfk \} ^{(0)}\big\}^{(2)}\otimes
\{ \epsilon_s^* \otimes \epsilon_s  \}^{(2)}\Big\}^{(0)} \Big\rangle
=
-\frac{1}{2\sqrt{15}}(\frac{1}{3} - |\epsilon \cdot \bfk_s|^2).
\end{eqnarray*}

This leads to Equation\eqref{e2e1av}.


\end{document}